\documentclass[11pt]{article}
\usepackage[dvips]{graphicx}
\usepackage{mathrsfs}
\usepackage{amsthm}  
\usepackage{amsmath}
\usepackage{amssymb}
\usepackage{amsfonts}
\usepackage{float}
\usepackage{yhmath}
\usepackage{wrapfig}
\usepackage[dvips]{color}
\usepackage[dvips]{colortbl}
\usepackage{cite}
\usepackage{array}
\usepackage{tabularx}
\usepackage[sectionbib]{chapterbib}
\usepackage{threeparttable}
\usepackage{chemarrow}
\usepackage[version=3]{mhchem}
\usepackage{framed}
\usepackage{ascmac}
\definecolor{shadecolor}{gray}{0.90}
\usepackage[small, bf, labelsep=quad]{caption}
\usepackage{cases}

\newcommand{\CF}{C_{\textit{\textsf{\hspace{-0.3mm}F}}}}
\newcommand{\NA}{N_{\textit{\textsf{\hspace{-0.3mm}Av}}}}

\newcommand{\I}{I}
\newcommand{\II}{I\hspace{-0.5mm}I}
\newcommand{\fw}{\langle f\rangle_{\hspace{-0.3mm}w}}
\newcommand{\gw}{\langle g\rangle_{\hspace{-0.3mm}w}}

\definecolor{mycolor1}{rgb}{1,1,0.7}
\definecolor{mycolor2}{rgb}{0.9,1,1}
\definecolor{mycolor3}{cmyk}{0,0,0,0.113}
\definecolor{mycolor4}{cmyk}{0.086,0,0,0}

\begin{document}

\begin{center}
{\huge On the Determination of Gel Points}
\end{center}
{\Large Correspondence between the Viscoelastic Theory and the Theory of Gelation}

\vspace*{5mm}
\begin{center}
\normalsize{Kazumi Suematsu\footnote{\, The author takes full responsibility for this article.}, Haruo Ogura$^{2}$, Seiichi Inayama$^{3}$, and Toshihiko Okamoto$^{4}$} \vspace*{2mm}\\
\normalsize{\setlength{\baselineskip}{12pt} 
$^{1}$ Institute of Mathematical Science\\[1mm]
Ohkadai 2-31-9, Yokkaichi, Mie 512-1216, JAPAN\\
E-Mail: suematsu@m3.cty-net.ne.jp, ksuematsu@icloud.com  Tel/Fax: +81 (0) 593 26 8052}\\[3mm]
$^{2}$ Kitasato University,\,\, $^{3}$ Keio University,\,\, $^{4}$ Tokyo University\\[10mm]
\end{center}

\hrule
\vspace{3mm}
\noindent
\textbf{\large Abstract}: A critical composition of cross-linked polysiloxanes observed by Scanlan and Winter is reinvestigated in comparison with the theory of gelation. We assume, based on the Scott findings, the geometric distribution for one of the monomers, divinyl-terminated poly(dimethylsiloxane). Calculation results show that the two theories are in near-consistency, supporting the Scanlan-Winter estimation based on the linear viscoelastic theory. On the other hand, there is a disturbing result that calculation using the mean molecular weight, $M_{n}$, leads to exact agreement between the two theories, suggesting that the distribution is in effect monodisperse, contrary to the assumed geometric one and also the observed polydispersity, $M_w/M_n=2.1$. Further experimental studies employing monodisperse monomers would be highly valuable to consolidate the bridge between these two fundamental theories.
\vspace{0mm}
\begin{flushleft}
\normalsize{}\\[3mm]
\end{flushleft}
\hrule
\vspace{8mm}
\setlength{\baselineskip}{14pt}

\vspace{0mm}
\section{Introduction}
A gel point is defined as a point where an infinitely large molecule emerges\cite{Flory}. Experimental determination of the gel point is a difficult task. The identification of the gel point still depends on the classical technique, the visual confirmation of the solidification of the reaction media, combined with the solubility test. Quite in contrast, the theoretical approach provides two promising ways to predict the gel point. One is the linear viscoelastic theory\cite{Tung, Winter}, which is based on the constitutive equation for viscoelastic media
\begin{equation}
	f(t)=\int_{-\infty}^t\hspace{-2mm}G(t-t')\frac{d \gamma(t')}{d t'}dt'\label{intro-1}
\end{equation}
where $f$ denotes applied force, $G$ relaxation modulus, $\gamma$ stress, and $t$ time. If a sinusoidal strain, $\gamma=\gamma_0\sin(\omega t')$, is applied to the system, and given the assumption of the power law, $G=S(t-t')^{-n}$, in the vicinity of gel points ($t'\rightarrow t=t_c$ from below), Eq. (\ref{intro-1}) can be rewritten in terms of the gamma function:
\begin{equation}
	f(t)=\gamma_0S\,\Gamma(1-n)\left\{\cos\left(\frac{n\pi}{2}\right)\sin(\omega t)+\sin\left(\frac{n\pi}{2}\right)\cos(\omega t)\right\}\,\omega^n\label{intro-2}
\end{equation}
Thus, the applied force can be expressed as a sum of a pure elastic component and a pure viscous component:
\begin{align}
	G'=&S\Gamma(1-n)\cos\left(\frac{n\pi}{2}\right)\,\omega^n\\
	G''=&S\Gamma(1-n)\sin\left(\frac{n\pi}{2}\right)\,\omega^n
\end{align}
While the initial observations by Tung and Dynes\cite{Tung} showed that various systems have the exponent $n=1/2$, the later experiments revealed that the “$\tan\left(\frac{n\pi}{2}\right)=constant$” is a more general law for gel point identification. Winter and coworkers\cite{Winter} put forth that gel points should be identified as the point at which $G''/G'=\tan\left(\frac{n\pi}{2}\right)$ is independent of $\omega$. Guided by this principle, Scanlan and Winter\cite{Winter} investigated the concentration dependence of gel points for a poly(dimethylsiloxane) system in a diluent, and observed that a critical dilution occurs at $\cong 5.2\, \text{wt}\%$.

An alternative  approach is the theory of gelation that leads us to the closed solution for the gel point, for instance, for the R$-$A$_f$ +  R$-$B$_g$ model\cite{Kazumi}:
\begin{equation}
p_{\text{A}c}=\sqrt{\frac{1}{s_0}}\left\{\frac{1-\displaystyle\frac{\sqrt{s_0}}{\NA}\sum\nolimits_{j=1}^\infty\varphi_{x}\left(1-\tfrac{1}{2j}\right)\,\gamma_{f}}{1-\displaystyle\frac{\sqrt{s_0}}{\NA}\sum\nolimits_{j=1}^\infty\varphi_{x}\,\gamma_{f}}\right\}\label{gel-5}
\end{equation}
where $p_{\text{A}c}$ denotes the extent of reaction of A functional units at the gel point, $s_0=(\fw-1)(\gw-1)/\kappa$, and $\kappa=\sum\nolimits_{j}g_{j}M_{\text{B}_{j}}/\sum\nolimits_{i}f_{i}M_{\text{A}_{i}}$. In Eq. (\ref{gel-5}), $\gamma_{f}=1/C_{f}$ is the reciprocal of A functional unit concentration that has the meaning of dilution, and $\varphi_{x}$ is the relative cyclization frequency of an $x$-chain to intermolecular reactions and has the form:
\begin{equation}
\varphi_x=\left(d/2\pi^{d/2}l_{s}^{\hspace{0.3mm}d}\right)\displaystyle\int_{0}^{d/2\nu_{\hspace{-0.3mm}x}}\hspace{-2mm}t^{\frac{d}{2}-1}e^{-t}dt \label{gel-6}
\end{equation}
where $\nu_{x}=\langle r_{x}^{2}\rangle/l_{s}^{2}=\CF\,\xi_{e}\, x$, and $l_{s}$ is the length of the cyclic bond.
As has been proven extensively, Eq. (\ref{gel-5}) reproduces well the experimental observations by Flory\cite{Flory}, Wile-Stockmayer\cite{Wile}, Gordon-Scantlebury\cite{Gordon}, Matejka\cite{Dusek}, Ilavsky-Dusek\cite{Dusek}, Budinski\cite{Budinski}, and many other workers.

Imposing the restraining condition, $p_{\text{A}c}\le 1$ for $\kappa\ge 1$ and $p_{\text{B}c}\le 1$ for $\kappa\le 1$, on Eq. (\ref{gel-5}), we have respectively
\begin{equation}
\gamma_{f}\le\hspace{2mm}
\begin{cases}
\hspace{3mm} \frac{1-1/\sqrt{s_0}}{\frac{1}{\NA}\displaystyle\sum\nolimits_{x}\left(-1+\sqrt{s_0}+1/2x\right)\varphi_{x}}\hspace{10mm}(\text{\I})\hspace{10mm} (\kappa\ge 1)\\[7mm]
\hspace{3mm} \frac{\kappa-1/\sqrt{s_0}}{\frac{1}{\NA}\displaystyle\sum\nolimits_{x}\left(-1+\kappa\sqrt{s_0}+1/2x\right)\varphi_{x}}\hspace{7mm}(\text{\II})\hspace{10mm} (\kappa\le 1)
\end{cases}\label{gel-7}
\end{equation}

We are going to apply the above inequalities to the gelation via the addition polymerization of divinyl-terminated poly(dimethylsiloxane) (DVPDS) and tetrakis(dimethylsiloxy)silane (TDS). According to literature\cite{Winter, Macosko}, the siloxane-prepolymer can be synthesized by the equilibrium polymerization of divinylsiloxane [CH$_2=$CH-Si(CH$_3$)$_2$-O-Si(CH$_3$)$_2$-CH$=$CH$_2$] and octamethylcyclotetrasiloxane (D4), so DVPDS is a mixture comprising various sizes of siloxane units (Fig. \ref{ReactionScheme}). In evaluating the production of rings, therefore, we must multiply the statistical weight, $\omega_n$, for the respective lengths of the prepolymers. The experiments\cite{Scott} have shown that the length distribution of the prepolymer is the geometric one: $P(n)=(1-p)p^{n-1}$ ($n =$ 1, 2, $\dots$) with the mean $\langle{n}\rangle=\sum_{n=1}^\infty nP(n)$, where $p$ is the bond probability of acyclic species alone. For the molecule sketched in Fig. \ref{ReactionScheme}, it follows that $p=\frac{n-1}{n}$. The end Si$-$O bond must be counted as an unreacted bond since it can neither disappear nor be created through the bond-exchange equilibria (c1):\\[-3mm]
\begin{shaded}
\begin{center}
	\hspace{-20mm}\ce{R3Si-[O-Si(CH3)2]_{n-1}-O-SiR3}\quad +\quad \ce{R3Si-[O-Si(CH3)2]_{m-1}-O-SiR3} \quad \ce{<=>}\\[3mm] \hspace{20mm}\qquad \ce{R3Si-[O-Si(CH3)2]_{n+q-1}-O-SiR3}\quad +\quad \ce{R3Si-[O-Si(CH3)2]_{m-q-1}-O-SiR3} \quad (c1)\\[5mm]
\end{center}
\vspace{-1mm}
\end{shaded}
\noindent where $n, m, q=1,2,3,\cdots$. 

The cross linking reaction is carried out by the reaction of the prepolymer (DVPDS) and TDS with the aid of platinum complexes, according to the chemical equation:
\begin{figure}[H]
\begin{center}
\includegraphics[width=17cm]{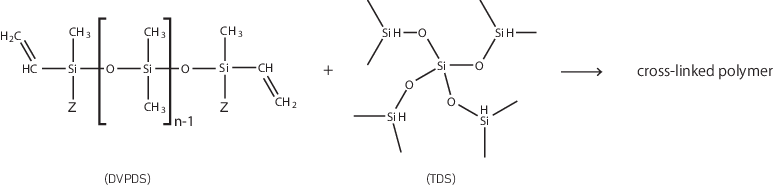}
\vspace{2mm}
\caption{\setlength{\baselineskip}{11pt}An AB-type polymerization of divinyl-terminated poly(dimethylsiloxane) and tetrakis(dimethyl-siloxy)silane. \textsf{Z}: CH$_3$ or Phenyl.}\label{ReactionScheme}
\end{center}
\vspace*{-4mm}
\end{figure}

\begin{figure}[H]
\begin{center}
\includegraphics[width=6cm]{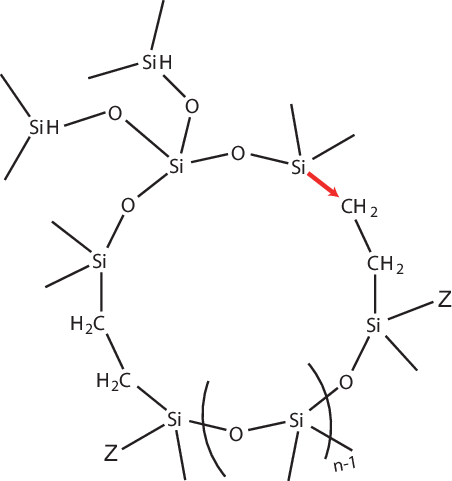}
\vspace{2mm}
\caption{\setlength{\baselineskip}{11pt}Macrocyclization of a linear AB monomer (n $=$ 1, 2, $\dots$). \textsf{Z}: CH$_3$ or Phenyl.}\label{smallestring}
\end{center}
\vspace*{-4mm}
\end{figure}

In Fig. \ref{smallestring}, the formation of a cyclic 1-mer is illustrated. The mean contour length of the chain monomer, the precursor of the ring monomer, is sufficiently long, so the Gaussian approximation is valid, given the condition of gel formation\cite{Kazumi}. The red Si$-$C bond denotes a cyclic bond with a length of $1.9\, \text{\AA}$. Then, following Fig. \ref{smallestring}, we can calculate the effective bond number:
\begin{equation}
\xi_{e, n}=\frac{1}{l_s^2}\left\{(2n'+4) l^2_\text{Si-O}+4\,l^2_\text{Si-C}+2\,l^2_\text{C-C}\right\}\label{gel-8}
\end{equation}
Here we apply $l_\text{Si-O}=1.64\,$\AA, $l_\text{Si-C}=1.9\,$\AA, and $l_\text{C-C}=1.37\,$\AA. Only, $n'$ has the geometric distribution. Since a chain length is evaluated as the total length of an $x$-mer, we must divide $n'$ by $x$, and hence $n'=n/x$. For this mixture, the quantity, $\nu_{x}$, is defined by  $\nu_{x}=\langle r_{x}^{2}\rangle/l_{s}^{2}=\CF\,(\xi_{e, n}\, x-1)$, where $l_s=l_\text{Si-C}=1.9\,$\AA \,, and $\CF=6.4$\, is the Flory characteristic constant\cite{Flory}. Using these data, we can calculate the relative cyclization frequency by the equations:
\begin{align}
\sum_{x=1}^\infty\frac{\varphi_x}{2x}=& \sum_{x=1}^\infty\sum_{n=kx}^\infty P_k(n,x) \frac{1}{2x}\left(d/2\pi^{d/2}l_{s}^{\hspace{0.3mm}d}\right)\int_{0}^{d/2\nu_{\hspace{-0.3mm}x}}\hspace{-0mm}t^{\frac{d}{2}-1}e^{-t}dt \label{gel-9} \\[3mm]
\sum_{x=1}^\infty\varphi_x=& \sum_{x=1}^\infty\sum_{n=kx}^\infty P_k(n,x) \left(d/2\pi^{d/2}l_{s}^{\hspace{0.3mm}d}\right)\int_{0}^{d/2\nu_{\hspace{-0.3mm}x}}\hspace{-2mm}t^{\frac{d}{2}-1}e^{-t}dt\label{gel-10}
\end{align}
$P_k(n,x)$ represents the probability that the total number of siloxane units in DVPDS constituting an $x$-chain, [a polymer with $x$ repeating units \ce{-(DVPDS-TDS)x-} in the backbone], is $n$. In Eqs. (\ref{gel-9}) and (\ref{gel-10}), the “$k$” means that lower molecular weight molecules ($n=1$ to $k-1$) have been devolatilized in vacuum in the process of the preparation of \text{DVPDS}. Thus, the prepolymer ($x=1$) comprises broad molecular species according to the distribution, $(1-p)p^{n-k}$, given the assumption that the geometric distribution is still obeyed after the devolatilization (Fig. \ref{MultiDistribution2}). 
 
Before proceeding with our discussion, we must examine the distribution function of the entire sequence of a chain $x$-mer, the precursor of a ring $x$-mer. 

\section{Generalized geometric Distribution and the Application}
\subsection{Scott's findings}
\begin{figure}[H]
	\vspace*{-2mm}
	\begin{minipage}[t][][s]{0.48\textwidth}
		\begin{center}
			\includegraphics[width=8.3 cm]{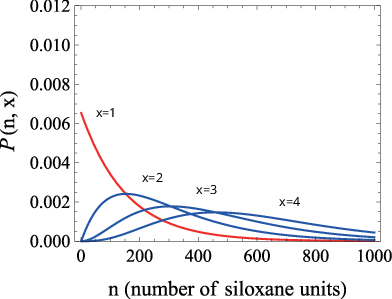}
		\end{center}
		\vspace*{-1mm}
		\caption{\setlength{\baselineskip}{11pt} Probability distribution function, $P(n,x)$, before devolatilization as a function of $n$ and $x$. $\langle n\rangle_n=153\,x$ ($p=\frac{152}{153}$).}\label{MultiDistribution1}
	\end{minipage}
	\hspace{5mm}
	\begin{minipage}[t]{0.48\textwidth}
		\begin{center}
			\includegraphics[width=8.3cm]{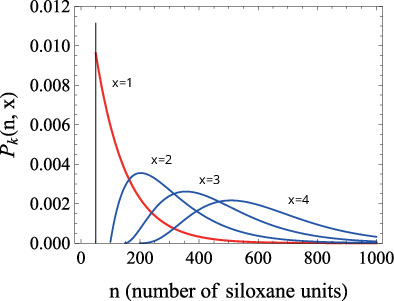}
		\end{center}
		\vspace*{-1mm}
		\caption{\setlength{\baselineskip}{11pt} Probability distribution function, $P_k(n,x)$, after devolatilization as a function of $k$, $n$ and $x$. In this example, $k=50$ and  $\langle n\rangle_n=153\,x$ ($p=\frac{103}{104}$).}\label{MultiDistribution2}
	\end{minipage}
\end{figure}

End-capped siloxane-prepolymers undergo the acid- or base-catalyzed equilibrium polymerization of the type (c1). Scott demonstrated\cite{Scott} that the products have the geometric distribution: $P(n, 1)=(1-p)p^{n-1}$. This formula applies to the formation of the smallest ring ($x=1$) shown in Fig. \ref{smallestring}. For the formation of higher rings, we need the extension of this formula. For this purpose, it is convenient to introduce the convolution; for instance, for a dimer chain ($x=2$), we have $P(n, 2)=\sum_{n_1=1}^{n-1}P(n_1,1)\cdot P(n-n_1,1)=(1-p)^2 p^{n-2}(n-1)$. In general, for an $x$-mer chain, \ce{-(DVPDS-TDS)x-}, comprising $x$ DVPDS, the precursor of an $x$-mer ring, we have the probability distribution function (PDF):
\begin{equation}
	P(n, x) =\hspace{-5mm}\sum_{\{n_1,n_2,\cdots,n_x\}}\hspace{-5mm}P(n_1,1)\cdot P(n_2,1)\cdots P(n_x,1)=\binom{n-1}{x-1} (1-p)^{x} p^{(n-x)}\hspace{1cm}(n\ge x) \label{gel-11}
\end{equation}
where $n_1+n_2+\cdots+n_x=n$. Eq. (\ref{gel-11}) satisfies the normalization condition: $\sum_{n=x}^{\infty}P(n,x)=1$, which can be proven by the Taylor expansion of $(1-p)^{-x}$ with respect to $p$. From this, we can derive the quantities: 
\vspace{0mm}
\begin{align}
&\text{number average}:\,\, \langle n\rangle_n=\sum_{n=x}^\infty nP(n,x)=\frac{x}{1-p}\label{gel-12}\\[-1mm]
&\text{weight average}:\,\, \langle n\rangle_w=\sum_{n=x}^\infty n^2P(n,x)/ \sum_{n=x}^\infty nP(n,x)=\frac{x+p}{1-p}\label{gel-13}\\[0mm]
&\text{polydispersity}:\,\, \langle n\rangle_w/\langle n\rangle_n = 1+\frac{p}{x}\label{gel-14}
\end{align}
The polydispersity varies from $1+p$ to 1 as the number of repeating units, $x$, increases from 1 to $\infty$. For $x\ge 2$, the transformation of the distribution occurs from the geometric distribution to the bell-shaped one (Fig. \ref{MultiDistribution1}). This is an example of the manifestation of the central limit theorem. 

The generalized PDF, $P_k(n,x)$, appearing in Eqs. (\ref{gel-9}) and (\ref{gel-10}) can be expressed in the form:
\begin{equation}
P_k(n,x)=\binom{n-(k-1)x-1}{x-1}(1-p)^xp^{(n-kx)}\hspace{1cm}(n\ge kx)\label{gel-15}
\end{equation}
for the polymers in which all of the short molecules less than $k$ have been removed, while the form of the PDF has remained the same. For $k=1$ (no devolatilization), Eq. (\ref{gel-15}) reduces to  Eq. (\ref{gel-11}). Eq. (\ref{gel-15}) satisfies the normalization condition: $\sum_{n=kx}^{\infty}P_k(n, x)=1$, and yields the basic quantities:
\vspace{0mm}
\begin{align}
	&\text{number average}:\,\, \langle n\rangle_n=\sum_{n=kx}^\infty nP_k(n,x)=\frac{[k(1-p)+p]x}{1-p}\label{gel-16}\\[-1mm]
	&\text{weight average}:\,\, \langle n\rangle_w=\sum_{n=kx}^\infty n^2P_k(n,x)/ \sum_{n=kx}^\infty nP_k(n,x)=\frac{p+[k(1-p)+p]^2x}{(1-p)[k(1-p)+p]}\label{gel-17}\\[0mm]
	&\text{polydispersity}:\,\, \langle n\rangle_w/\langle n\rangle_n = 1+\frac{p}{[k(1-p)+p]^2x}\label{gel-18}
\end{align}
For $k=1$, these recover Eqs. (\ref{gel-12})-(\ref{gel-14}). Fig. \ref{MultiDistribution2} shows an example of the devolatilized prepolymers of $k=50$, assuming $\langle n\rangle_n=153\,x$. 
\subsection{Valles-Macosko experiments}\label{Valles-Macosko}
The aim of this section is to evaluate the “$k$” value appearing in Eqs. (\ref{gel-9}) and (\ref{gel-10}). This value should be estimated by point-by-point comparison of the observed values with theoretical gel points that can be calculated as a function of varying $k$ by using the gelation equation (\ref{gel-5}). In the preparation of polysiloxane networks, the monomer, DVPDS (a mixture of various sizes of linear siloxanes: X = Phenyl), was synthesized by the equilibrium polymerization mentioned above, and was devolatilized to remove the low molecular weight portion together with cyclic species that are detrimental to final products. Valles and Macosko\cite{Macosko} report that cyclic dimethylsiloxanes D4 through D20 remaining in the reaction products were almost removed after one week of treatment at $100^{\circ}$C in vacuum. On one hand, according to the literature\cite{Hunter}, a cyclic siloxane and a linear siloxane have nearly the same boiling point for the same molecular weight (see Fig. \ref{BP-MW}). So, $k\ge 20$ is assured\cite{Brown}. The extent of reaction, $p$, can be estimated by solving Eq. (\ref{gel-15}) for an arbitrary $x$ and taking the chemical constituents of DVPDS into consideration, as $\langle n\rangle_n=\sum_{n=kx}^{\infty}nP_k(n,x)\cong 153x$, with the result $p=\frac{153-k}{154-k}$. Then, using Eqs. (\ref{gel-9}) and (\ref{gel-10}), the production of rings are calculated. It is seen that the mean value method estimates a lesser number of rings, in agreement with the previous findings\cite{Kazumi}. 

\begin{center}
	\begin{threeparttable}[h]
		\caption{Calculation results for the relative cyclization frequencies for $\kappa\doteq 1$($d=3$).}\label{Table-1}
		\begin{tabular}{c c c c c}
			\hline\\[-3mm]
			\hspace{0mm}$k$ values\tnote{\,a} & \hspace{4mm}$\displaystyle\sum\nolimits_{x=1}^\infty\frac{\varphi_x}{2x}$\,\, $(moles/L)$ & \hspace{3mm}$\displaystyle\sum\nolimits_{x=1}^\infty\varphi_x$\,\, $(moles/L)$ & \hspace{3mm}$p_c$(ex.)\tnote{\,b} &  \hspace{3mm}$p_c$(theor.) \\[3mm]
			\hline\\[-1.5mm] 
			\hspace{0mm}$k=20$ & \hspace{5mm}0.002458 & \hspace{4mm}0.006947 & \hspace{5mm}  & \hspace{5mm}0.591-0.593
			 \\[2mm]
			\hspace{0mm}$k=30$ & \hspace{5mm}0.001987 & \hspace{4mm}0.005897 & \hspace{5mm} & \hspace{5mm}0.587-0.590\\[2mm]
		    \raisebox{-1mm}{$k=40$} & \raisebox{-1mm}{\hspace{5mm}0.001698} & \raisebox{-1mm}{\hspace{4mm}0.005237} & \hspace{5mm} & \raisebox{-1mm}{\hspace{5mm}0.586-0.588} \\[2mm]
			\rowcolor[rgb]{1,1,0.6}\raisebox{-1mm}{\hspace{0mm}$k=50$} & \raisebox{-1mm}{\hspace{5mm}0.001501} & \raisebox{-1mm}{\hspace{4mm}0.004779} & \hspace{5mm} & \raisebox{-1mm}{\hspace{5mm}0.584-0.587}
			 \\[2mm]
			\rowcolor[rgb]{1,1,0.6}\hspace{-1mm}	\raisebox{-1mm}{\hspace{0mm}$k=60$} & \raisebox{-1mm}{\hspace{5mm}0.001357} & \raisebox{-1mm}{\hspace{4mm}0.004441} & \raisebox{-1mm}{\hspace{5mm}} & \raisebox{-1mm}{\hspace{5mm}0.583-0.586}
			\\[2mm]
				\rowcolor[rgb]{0.6,0.8,1}\hspace{0mm}\raisebox{-2mm}{mean value method}\tnote{\,c} & \raisebox{-2mm}{\hspace{5mm}0.000919} & \raisebox{-2mm}{\hspace{4mm}0.003568} & \hspace{5mm} & \raisebox{-2mm}{\hspace{5mm}0.581-0.583} \\[3mm]
			\hline\\[-5mm]
		\end{tabular}
		
		\begin{tablenotes}
			\item a. $k$ means the smallest molecule in the prepolymer, \text{DVPDS}, a mixture of various sizes of end-capped linear siloxanes.
			\item b. 0.581-0.588 by Valles and Macosko (their $k$ value is unknown).
			\item c. In Eqs. (\ref{gel-9}) and (\ref{gel-10}), $\sum_{n=kx}^\infty P_k(n, x)\equiv 1$, and in Eq. (\ref{gel-8}), $n'\equiv 153$.
		\end{tablenotes}
	\end{threeparttable}
\end{center}

 In Table \ref{Table-1}, the results are summarized for varying $k$, which vary depending on the molar ratios of DVPDS to TDS ($\kappa=1/0.999$ to $1/1.008$). Comparing those with the observed values, $p_c(ex)=0.581-0.588$, we find that the $k$ value is around $50\sim 60$. This raises a question whether the components possessing such high molecular weight (MW) of $3500\sim4500$ can really be devolatilized by the treatment of 1 week at $100^{\circ}$C in vacuum, while the experiments\cite{Macosko} have suggested that the cyclic siloxane D20 ($\text{MW}=1480$) can be devolatilized under the same condition,  as mentioned above.
 
\subsection{Scanlan-Winter experiments}
Based on the observations by Scott, and those by Valles and Macosko, we proceed to the main theme of this paper. Scanlan and Winter\cite{Winter} investigated the concentration dependence of gel points  utilizing the same cross-linking system as discussed above. According to their paper, DVPDS was commercially supplied by HULS Petrarch (the synthetic method is not disclosed, so the distribution function is unknown) and devolatilized before use at $140^{\circ}$C for 14h in vacuum. For their DVPDS, we assume the probability distribution function (\ref{gel-15}). The validity of this assumption can be supported by the observed dispersion, $\langle n\rangle_w/\langle n\rangle_n=2.1$. The number average molecular weight was $\cong 10300$, and $\langle n\rangle_n\cong 140$. By the observations by Valles and Macosko together with the calculation results in Table \ref{Table-1}, we estimate that $k=50\sim 60$, from which we calculate the extent of reaction of the prepolymer, by solving Eq. (\ref{gel-15}), as $\langle n\rangle_n=\sum_{n=kx}^{\infty}nP_k(n,x)\cong 140x$, for each $k$ value (see table notes in Table \ref{Table-2}). Then, the production of rings is readily calculated by using Eqs. (\ref{gel-9}) and (\ref{gel-10}).

\begin{center}
  \begin{threeparttable}[h]
    \caption{Estimation of critical dilution points, $\gamma_{f_{c}}$.}\label{Table-2}
  \begin{tabular}{l c c c}
\hline\\[-3mm]
\hspace{3mm}Calculation methods & \hspace{3mm}$\displaystyle\sum\nolimits_{x=1}^\infty\frac{\varphi_x}{2x}$\,\, $(moles/L)$ & \hspace{3mm}$\displaystyle\sum\nolimits_{x=1}^\infty\varphi_x$\,\, $(moles/L)$ & \hspace{3mm}$\gamma_{f_{c}}$\tnote{\,c}\hspace{4mm}($\kappa=1.7$)\\[3mm]
\hline\\[-1.5mm]
\hspace{3mm}$P_k(n,x)$, $k=50$ & \hspace{3mm}0.00162\tnote{\,a} & \hspace{3mm}0.00523\tnote{\,a} & \hspace{3mm}74.02\\[2mm]
\hspace{3mm}$P_k(n,x)$, $k=60$ & \hspace{3mm}0.00147\tnote{\,b} & \hspace{3mm}0.00487\tnote{\,b} & \hspace{3mm}80.57\\[2mm]
\hspace{3mm}Mean value method\tnote{\,d} & \hspace{3mm}0.00104 & \hspace{3mm}0.00406 & \hspace{3mm}104.0\\[2mm]
\hline\\[-6mm]
   \end{tabular}
    \vspace*{2mm} 
   \begin{tablenotes}
     \item a. Calculated by Eqs. (\ref{gel-9})-(\ref{gel-10}) with $p=90/91$.
     \item b. Calculated by Eqs. (\ref{gel-9})-(\ref{gel-10}) with $p=80/81$.
     \item c. Calculated by Eq. (\ref{gel-7}-\I).
     \item d. Calculated by  Eqs. (\ref{gel-9}) and (\ref{gel-10}), but $\sum_{n=kx}^\infty P_k(n, x)\equiv 1$ and $ n'\cong 140$.
   \end{tablenotes}
  \end{threeparttable}
\end{center}

\begin{figure}[H]
\begin{center}
\includegraphics[width=7.5cm]{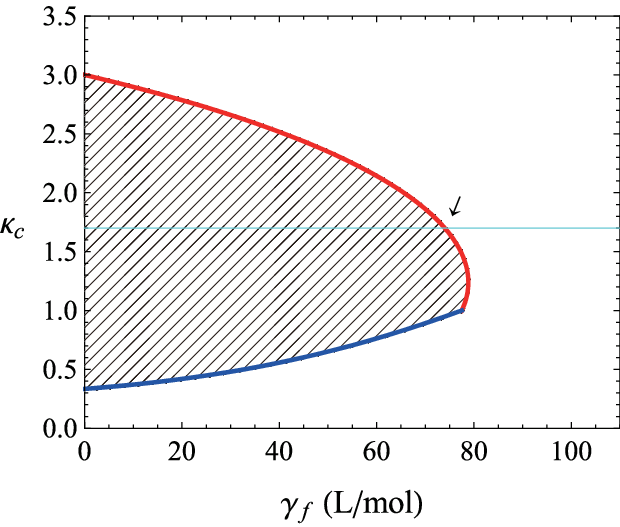}
\vspace{-2mm}
\caption{\setlength{\baselineskip}{11pt}Critical molar ratios against the reciprocal of the A-functional units concentration $\left([\text{C}=\text{C}]\right)$. Calculated by using the data of $k=50$ in the polymerization of divinyl-terminated poly(dimethylsiloxane) and tetrakis(dimethylsiloxy)silane. The shaded region represents the area where gelation is possible.}\label{Fig-4}
\end{center}
\vspace*{-4mm}
\end{figure}
Substituting the numerical values thus obtained, together with $f=2$ and $g=4$, into Eq. (\ref{gel-7}), we can simulate critical molar ratios, $\kappa_c$, as a function of the dilution, $\gamma_f$, of A functional units. In Fig. \ref{Fig-4}, an example of a simulation for $k=50$ is shown: the red solid line is by Eq. (\ref{gel-7}-\I)\, and the blue solid line is by Eq. (\ref{gel-7}-\II). The inner shaded region represents the area where gelation is possible. The arrow in Fig. \ref{Fig-4} indicates the critical dilution for $\kappa=1.7$ that corresponds to the Scanlan-Winter experiments. All the results are shown in Table \ref{Table-2}.

\section{Correspondence with the Theory of Gelation}
Scanlan and Winter carried out the above addition polymerization in 1991, while the concept of critical dilution was introduced in 2002\cite{Kazumi}. So, at the time of the Scanlan and Winter experiments, the phenomenon of critical dilution was not known. The classical, critical molar ratio derived as $1/(f-1)(g-1) \le \kappa_{classical} \le (f-1)(g-1)$ was, of course, known. However, what we are discussing in this paper is the critical dilution that occurs through the cyclization effects, an event in the interval of the classical critical molar ratio. So, contrary to the critical molar ratio, the critical dilution depends strongly on the system concentration. The occurrence of such critical dilution has been fully confirmed through comparison with numerous experimental observations\cite{Wile, Dusek, Budinski, Kazumi}.  

The central objective of this paper is to identify the critical dilution point for the above-mentioned system in light of the theory of gelation and compare the result with the observed value by Scanlan and Winter. In this way we want to confirm the consistency between the theory of gelation and the linear viscoelastic theory. According to the linear viscoelastic theory, gelation should occur at the point where the storage and the loss moduli have the same frequency dependence, namely, $G',\,G''\propto \omega^n$. Scanlan and Winter observed that the critical composition for the system with the stoichiometry ($\kappa=1.7$) occurs at $5.2 \,\text{wt}\%$ prepolymer.

As Table \ref{Table-2} shows, the theory of gelation predicts the critical dilution point, $\gamma_{f_c}=74.02\,\,L/mole$ (for the geometric distribution of $k=50$) and $\gamma_{f_c}=80.57\,\,L/mole$ ($k=60$). The prepolymer has the mean molecular weight of $\text{MW}\cong 10300$ and the specific gravity $\rho\cong 0.965$. Since $f=2$, we find immediately the critical weight fraction of the prepolymer:
\begin{equation}
w_c\cong\begin{cases}
\displaystyle\frac{10300}{2\cdot 74.02\cdot 0.965}=72.5\,\,\text{g/Kg}\hspace{5mm}(k=50)\\[7mm]
\displaystyle\frac{10300}{2\cdot 80.57\cdot 0.965}=66.6\,\,\text{g/Kg}\hspace{5mm}(k=60)
\end{cases}\label{gel-19}
\end{equation} 
which are $\cong 7.3\,\text{wt}\%$ and $\cong 6.7\,\text{wt}\%$ respectively. The results are in good conformity with the Scanlan-Winter observations ($\cong 5.2 \,\text{wt}\%$). On the other hand, the mean value method gives an attractive answer, $\frac{10300}{2\cdot 104\cdot 0.965}=51.6\,\,\text{g/Kg}\, (\cong 5.2 \,\text{wt}\%)$, exactly equal to the observed value (Table \ref {Table-3}). However, the mean value method ignores the polydispersity of the prepolymers and the presence of short molecules, while cyclization is susceptible, particularly, to short molecules and enhanced by dilution. For this reason, the remarkable agreement with the observed value must be carefully reinvestigated by using pure monomers.

It must be emphasized that the results in Table \ref{Table-3} have been deduced under the assumption that linear polysiloxanes with $\text{MW}=3500\sim 4500$ are volatile and can therefore be removed by heat treatment in vacuum (the $k=50$ and $60$ cases). Stretching this line, if we apply larger $k$ values, for instance, $k=65, \cdots$, we will be able to obtain a better agreement with the observed value, which, however, forces us to extend a dubious volatility hypothesis to still higher molecular weight polysiloxanes. What is the molecular weight limit for the volatility? A hint may exist in the paper by Valles-Macosko, who suggested that the cyclic siloxane, D20 ($\text{MW} = 1480$), can be removed at $100^\circ$C in vacuum. Hunter, Warrick, Hyde, and Currie\cite{Hunter} showed that cyclic siloxanes and linear ones have almost the same boiling points (see Fig. \ref{BP-MW} in Appendix \ref{Appendix A}; note that polydimethylsiloxanes are far more volatile than corresponding polyhydrocarbons!); hence, $k\geq 1500$ is ensured. If the volatility hypothesis of the polymers having $\text{MW}=3500\sim 4500$ is false, we must employ smaller $k$'s. Then the difference between the prediction of the theory of gelation and that of the linear viscoelastic theory will expand. Another problem exists, namely, the problem that the experimental distribution of the prepolymers after the devolatilization has not been disclosed \cite{Macosko, Winter}. If those have the bell-shaped distribution (Gaussian, Poisson), the estimation by the mean value method takes on importance\cite{Kazumi}. Some literature\cite{Fuchise} reports that a linear poly(dimethylsiloxane) can be synthesized by means of anionic living polymerizations to yield a narrow polydispersity ($\langle n\rangle_w/\langle n\rangle_n\cong 1$). If DVPDS was in fact synthesized by the living polymerization, the prepolymer is expected to have the Poisson distribution. Then the mean value method is exact. Unfortunately, such a conjecture might be incorrect, because experiments have shown that the prepolymers have the polydispersity of $\langle n\rangle_w/\langle n\rangle_n\cong 2.1-2.2$\cite{Macosko, Winter} consistent with the geometric distribution rather than the Poisson one.

\begin{center}
	\begin{threeparttable}[h]
		\caption{Comparison between the linear viscoelastic theory and the theory of gelation: for gelation expriments in the polymerization of divinyl-terminated poly(dimethylsiloxane) and tetrakis(dimethylsiloxy)silane at $\kappa=1.7$.}\label{Table-3}
		\begin{tabular}{l c c c}
			\hline\\[-3mm]
			\hspace{3mm}Calculation methods & \hspace{7mm}$\gamma_{f_{c}}$\,$(L/mole)$\hspace{4mm}& \hspace{7mm}Critical weight \% of prepolymer\\[3mm]
			\hline\\[-1.5mm]
			\hspace{3mm}Viscoelastic theory\tnote{\,a} & \hspace{7mm} & \hspace{6mm}5.2 & \hspace{7mm}\\[2mm]
			\hspace{3mm}Theory of gelation ($k=50$)\tnote{\,b} & \hspace{7mm}74.02 & \hspace{6mm}$7.3$ &\hspace{7mm}\\[1.5mm]
			\hspace{3mm}Theory of gelation ($k=60$)\tnote{\,b} & \hspace{7mm}80.57 & \hspace{6mm}$6.7$ &\hspace{7mm}\\[1.5mm]
			\hspace{3mm}Theory of gelation (mean value method)\tnote{\,c} & \hspace{7mm}104.0 & \hspace{6mm}$5.2$ &\hspace{7mm}\\[1.5mm]
			\hline\\[-6mm]
		\end{tabular}
		\vspace*{2mm}
		\begin{tablenotes}
			\item a. by Scanlan and Winter\cite{Winter}.
			\item b, c. This work.
		\end{tablenotes}
	\end{threeparttable}
\end{center}

\section{Concluding Remarks}
Through this work, we have laid the assumption that the devolatilization is clear-cut at $n=k$, so that the molecules having $n\ge k$ obey the geometric distribution (Fig. \ref{MultiDistribution2}). If linear polysiloxanes having $\text{MW}=3500\sim4500$ are in fact volatile in vacuum, agreement between the linear viscoelastic theory and the theory of gelation is satisfactory.

 On the other hand, there is a disturbing result that the mean value calculation gives a remarkable agreement with the observed value. If that is true, it follows that the distribution is in effect monodispersed, contrary to the observed polydispersity, $\langle n\rangle_w/\langle n\rangle_n\cong 2.1-2.2$ \cite{Macosko,  Winter}.
 
To summarize, the following different conclusions are possible:
\begin{enumerate}
	\item The prepolymer, DVPDS, obeys the geometric distribution, and polysiloxanes having molecular weights of 3500-4500 are volatile. Given these assumptions, the theory of gelation and the linear viscoelastic theory are in good consistency within experimental errors. The two theories complement each other.
    \item The prepolymer, DVPDS, obeys the bell-shaped distribution, so that the mean value method is a good approximation. Given this assumption, the theory of gelation and the linear viscoelastic theory predict precisely the same gel point.
\end{enumerate}
In either case, the viscoelastic theory and the theory of gelation predict the critical dilution point with good approximation. However, the use of the mixture of prepolymers obscures the precise instant of the gelation and, hence, is not suitable for the rigorous assessment of the physical equivalence of the ``$\tan\left(\frac{n\pi}{2}\right)=const.$ theorem'' and the theory of gelation. To resolve this question, future experiments with well-defined precursor chains\cite{Flory, Wile, Gordon, Ross-Murphy, Kazumi} will provide deeper clarity on this consistency.

\appendix
\section{Boiling Point Difference between Hydrocarbons and Dimethylsiloxanes}\label{Appendix A}
\begin{figure}[H]
	\begin{center}
		\includegraphics[width=8cm]{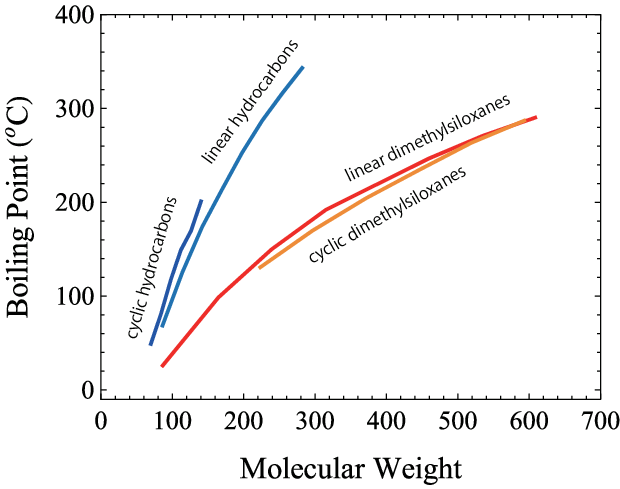}
		\vspace{2mm}
		\caption{\setlength{\baselineskip}{11pt}Molecular Weight Dependence of Boiling Points.}\label{BP-MW}
	\end{center}
\end{figure}

As one can see from Fig. \ref{BP-MW}, dimethylsiloxane compounds (chains and rings) are more volatile than corresponding hydrocarbons. The data are due to Hunter, Warrick, Hyde, and Currie\cite{Hunter}.

\vspace{1cm}

\end{document}